\documentclass[%
 reprint,
 amsmath,amssymb,
 aps,
prb,
]{revtex4-1}
\usepackage[utf8]{inputenc}
\usepackage{graphicx}
\usepackage{dcolumn}
\usepackage{bm}
\usepackage{hyperref}
\usepackage{amsmath}
\usepackage{amssymb}
\usepackage{pdfpages}
\makeatletter
\AtBeginDocument{\let\LS@rot\@undefined}
\makeatother

\begin{document}
\title{Oscillating magnetoresistance in graphene \textit{p}-\textit{n} junctions at intermediate magnetic fields }

\author{Hiske Overweg}

\author{Hannah Eggimann}
\affiliation{%
Solid State Physics Laboratory, ETH Zürich,~CH-8093~Zürich, Switzerland}
\author{Ming-Hao Liu}
\affiliation{%
Institut für Theoretische Physik, Universität Regensburg, D-93040 Regensburg, Germany}
\affiliation{%
Department of Physics, National Cheng Kung University, Tainan 70101, Taiwan}

\author{Anastasia Varlet}
\author{Marius Eich}
\author{Pauline Simonet}
\author{Yongjin Lee}
\affiliation{%
Solid State Physics Laboratory, ETH Zürich,~CH-8093~Zürich, Switzerland}

\author{Kenji Watanabe}
\author{Takashi Taniguchi}
\affiliation{Advanced Materials Laboratory, National Institute for Material Science, 1-1 Namiki, Tsukuba 305-0044, Japan}
\author{Klaus Richter}
\affiliation{%
Institut für Theoretische Physik, Universität Regensburg, D-93040 Regensburg, Germany}
\author{Vladimir I. Fal'ko}
\affiliation{%
National Graphene Institute, University of Manchester,
Manchester M13 9PL, UK}
\author{Klaus Ensslin}
\author{Thomas Ihn}
\affiliation{%
Solid State Physics Laboratory, ETH Zürich,~CH-8093~Zürich, Switzerland}
 \email{overwegh@phys.ethz.ch}

\date{April 6, 2017}

\begin{abstract}
We report on the observation of magnetoresistance oscillations in graphene \textit{p}-\textit{n} junctions. The oscillations have been observed for six samples, consisting of single-layer and bilayer graphene, and persist up to temperatures of 30~K, where standard Shubnikov-de Haas oscillations are no longer discernible. The oscillatory magnetoresistance can be reproduced by tight-binding simulations. We attribute this phenomenon to the modulated densities of states in the \textit{n}- and \textit{p}- regions. 
\end{abstract}
\maketitle

\textit{p}-\textit{n} junctions are among the basic building blocks of any electronic circuit. The ambipolar nature of graphene provides a flexible way to induce \textit{p}-\textit{n} junctions by electrostatic gating. This offers the opportunity to tune the charge carrier densities in the \textit{n}- and \textit{p}-doped regions independently. The potential gradient across a \textit{p}-\textit{n} interface depends on the thickness of the involved insulators and can also be modified by appropriate gate voltages. Due to the high electronic quality of present day graphene devices a number of transport phenomena in \textit{pnp} or \textit{npn} junctions have been reported, such as ballistic Fabry-Pérot oscillations \cite{Grushina2013,Rickhaus2013,Varlet2014} and so-called snake states \cite{Taychatanapat2015,Rickhaus}, both of which depend on characteristic length scales of the sample. 

Here we report on the discovery of yet another kind of oscillation, which does not depend on any such length scale. The oscillations occur in the bipolar regime, in the magnetic field range where Shubnikov-de Haas oscillations are observed in the unipolar regime. These novel oscillations in the bipolar regime are governed by the unique condition that the distance between two resistance minima (or maxima) in gate voltage space is given by a constant filling factor difference of $\Delta \nu =~$8. The features are remarkably robust: they occur in samples with one and two \textit{p}-\textit{n} interfaces; in single and bilayer graphene; up to temperatures of 30 K (where Shubnikov-de Haas oscillations have long disappeared); over a large density range; for interface lengths ranging from 1~$\mu$m to 3~$\mu$m and in both \textit{pnp} and \textit{npn} regimes. The oscillations have been observed in a magnetic field range of $B = 0.4$~T up to $B = 1.4$~T. Their periodicity does not match the periodicity of the aforementioned snake states. In this paper we address this phenomenon and suggest a model which can qualitatively explain the oscillations.

\begin{figure}
\centering
\includegraphics[width=0.5\textwidth]{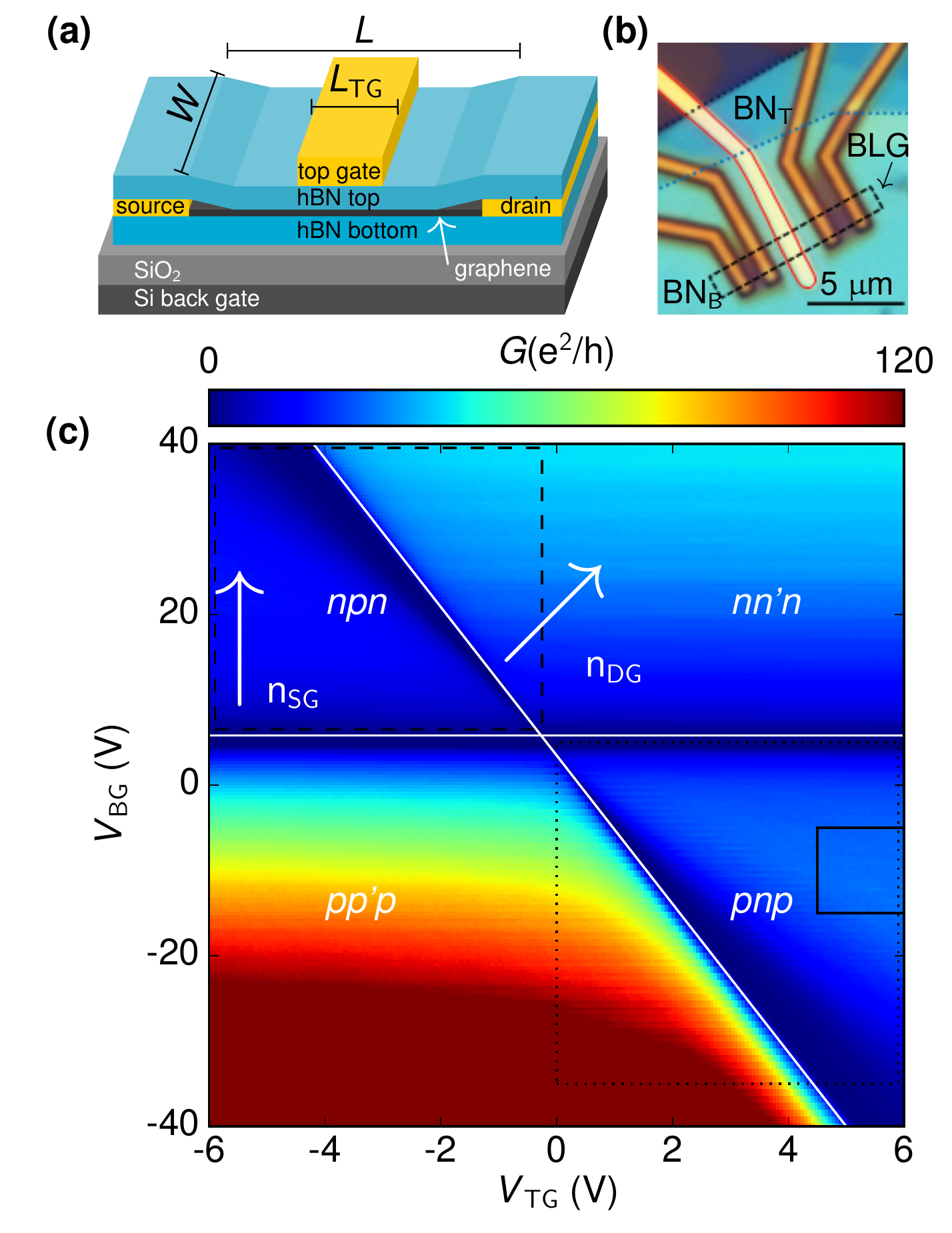}
\caption{Characterization of the device. (a) Schematic of the device: a bilayer graphene flake is encapsulated between h-BN layers. It is contacted by Au contacts and a Au top gate is patterned on top, which defines the dual-gated region. (b) Optical microscope image of the sample. The four contacts, of which only the inner ones were used, appear orange. The top gate is outlined by a red curve. (c) Conductance of the sample at $B = 0$~T, $T = 1.7$~K. Four regions of different polarities are indicated. A zoom of the transconductance in the boxed region with a solid line is shown in Fig.~\ref{fig:FP}. The dashed (dotted) box indicates the gate voltage range in which Figs.~\ref{fig:mapping}a,b (d,e) were measured.}
\label{fig:1}
\end{figure}

Measurements were performed on six samples in total, which all consist of a graphene flake encapsulated between two hexagonal boron nitride (h-BN) flakes on a Si/SiO$_2$ substrate. They all show similar behavior. This paper focuses on measurements performed on one sample (sample A), with the device geometry sketched in Fig.~\ref{fig:1}a. Specifications of the other five samples are summarized in table \ref{tab:sizes}.
\begin{table}
\resizebox{0.5 \textwidth}{!}{%
\begin{tabular}{|c||c|c|c|c|c|c|}
\hline 
\rule[-1ex]{0pt}{2.5ex} \textbf{sample name} & \textbf{A} & \textbf{B} & \textbf{C} & \textbf{D} & \textbf{E} & \textbf{F} \\ 
\hline 
\rule[-1ex]{0pt}{2.5ex} \textbf{sample width}  \textit{W} ($\mu$m) & 1.3 & 1.4 & 1.1 & 0.9 & 3 & 1.2 \\ 
\hline 
\rule[-1ex]{0pt}{2.5ex} \textbf{sample length} \textit{L} ($\mu$m) & 3.0 & 1.4 & 1.0 & 2.3 & 3 & 2.8  \\ 
\hline 
\rule[-1ex]{0pt}{2.5ex} \textbf{top gate length} \textit{L}$_\mathrm{TG}$ ($\mu$m) & 1.1 & 0.7 & 0.55 & 1.2 & 1.0 & 1.0  \\ 
\hline 
\rule[-1ex]{0pt}{2.5ex} \textbf{distance to top gate} (nm) & 23 & 44 & 28 & 57 & 35 & 25\\ 
\hline 
\rule[-1ex]{0pt}{2.5ex} \textbf{number of graphene layers} & 2 & 1 & 2 & 2 & 2 & 2 \\ 
\hline 
\rule[-1ex]{0pt}{2.5ex} \textbf{junction type} & \textit{npn} & \textit{pn} & \textit{npn} & \textit{pn} & \textit{npn} & \textit{npn} \\ 
\hline
 
\end{tabular}} 
\caption{Characteristics of samples A-F}
\label{tab:sizes}
\end{table}
The bilayer graphene (BLG) flake was encapsulated with the dry transfer technique described in Ref.~\citenum{Dean2010}. A top gate was evaporated on the middle part of the sample, which divides the device into two outer regions, only gated by the back gate (single-gated regions), and the dual-gated middle region. The other five samples were made with the more recent van der Waals pick-up technique.\cite{Wang2013} Unless stated otherwise, the measurements were performed at 1.7 K. An AC voltage bias of 50 $\mu$V was applied symmetrically between the Ohmic contacts (`source' and `drain' in Fig.~\ref{fig:1}a, inner contacts in Fig.~\ref{fig:1}b) and the current between the same contacts was measured. The transconductance $dG/dV_\mathrm{TG}$ was measured by applying an AC modulation voltage of 20 mV  to the top gate.

\begin{figure}
\centering
\includegraphics[width=0.5\textwidth]{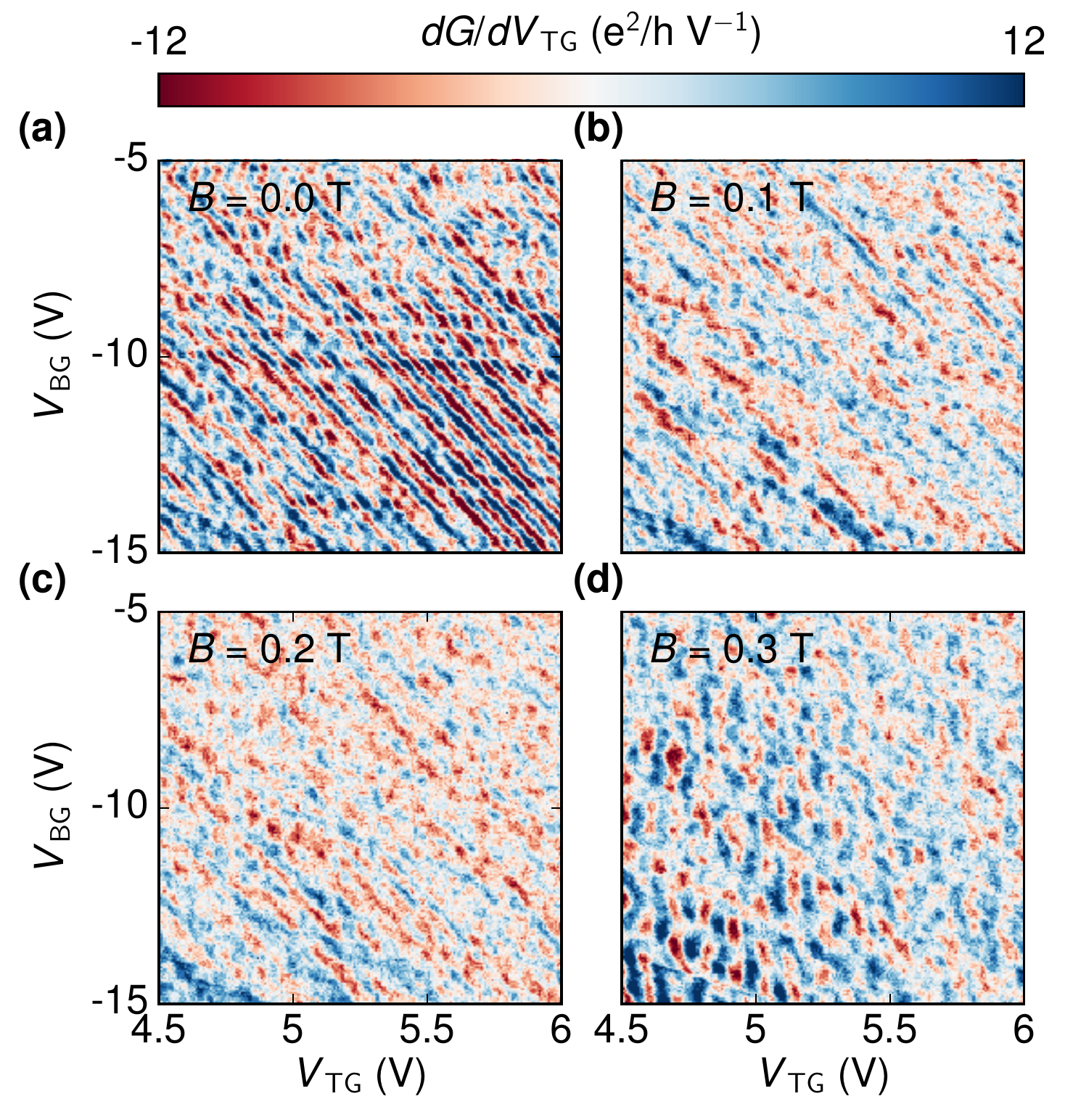}
\caption{Disappearance of Fabry-Pérot oscillations with increasing magnetic field. The measurement was taken in the boxed region with solid lines in Fig.~\ref{fig:1}c. At $B = 0$~T (a) the transconductance shows clear Fabry-Pérot oscillations. They disappear in a magnetic field of $B \gtrsim 0.1$~T (b--d).}
\label{fig:FP}
\end{figure}

Figure~\ref{fig:1}c shows the conductance as a function of top gate voltage $V_\mathrm{TG}$ and back gate voltage $V_\mathrm{BG}$. Charge neutrality of the single-gated regions shows up as a horizontal line of low conductance and is marked by a white line. The diagonal line of low conductance corresponds to charge neutrality of the dual-gated region. The slope of this line is given by the capacitance ratio of the top and back gate. Together these lines divide the map into four regions with different combinations of carrier types: two with the same polarities in the single- and dual-gated regions (\textit{pp'p} and \textit{nn'n}) and two with different polarities (\textit{npn} and \textit{pnp}). The conductance in the latter regions shows a modulation which is more clearly visible in the transconductance (see Fig.~\ref{fig:FP}a). The oscillatory conductance is caused by Fabry-Pérot interference of charge carriers travelling back and forth in the region of the sample underneath the top gate. Their periodicity yields a cavity length $L_\mathrm{TG} = 1.1~\mu$m, which is in agreement with the lithographic length of the top gate. The Fabry-Pérot oscillations were studied in more detail in Ref.~\citenum{Varlet2014}, which revealed the ballistic nature of transport in the dual-gated region.

\begin{figure}
\centering
\includegraphics[width=0.5\textwidth]{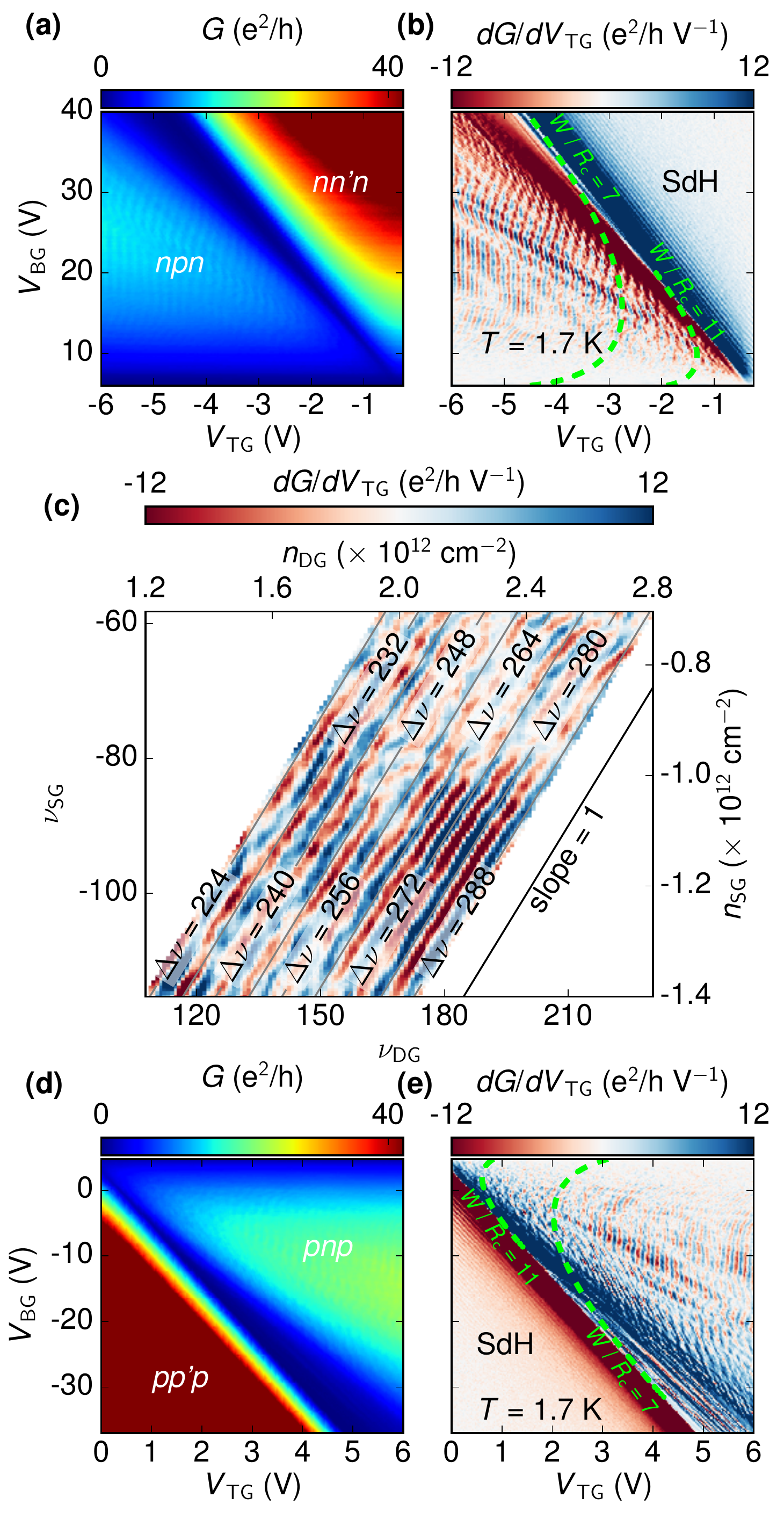}
\caption{Magnetotransport at $B = 0.5$~T. (a) Conductance of the sample at 0.5~T, showing an oscillatory pattern in the \textit{npn} regime. The measurement was taken in the dashed boxed region of Fig.~\ref{fig:1}c. (b) The oscillatory pattern in the \textit{npn} regime is more clearly visible in the transconductance. Green dashed lines indicate the pattern expected for snake states. In the \textit{nn'n} regime some faint lines can be distinguished, following the slope of the charge neutrality line of the dual gated region. These are Shubnikov-de Haas oscillations. (c) Transconductance at $B = 0.5$~T in the \textit{pnp} regime as a function of charge carrier density (and filling factor) in the single- and dual-gated region. The oscillatory pattern follows the indicated line of slope one and can therefore be described by lines of constant filling factor difference $\Delta \nu = \nu_\mathrm{DG}-\nu_\mathrm{SG}$. (d),(e) Same as (a),(b), but with opposite charge carrier polarities. The oscillations are essentially particle-hole symmetric.}
\label{fig:mapping}
\end{figure}

The Fabry-Pérot oscillations disappear in a magnetic field of $B \gtrsim 100$ mT (see Fig.~\ref{fig:FP}b-d). Yet at magnetic fields of $B = 0.4$~T a new oscillatory pattern appears in the \textit{npn} and \textit{pnp} regime. This can be seen in the conductance and transconductance maps recorded at $B = 0.5$~T, shown in Figs.~\ref{fig:mapping}a,b,d,e. The oscillations follow neither the horizontal slope of features taking place in the single-gated region, nor the diagonal slope of the dual-gated region. They are therefore expected to occur at the interface between the \textit{p}- and \textit{n}-doped regions. This was confirmed by measurements on sample D, which had two contacts in the single-gated region and two contacts in the dual-gated region. For this sample, only the conductance along paths involving the interface shows oscillations (see Supporting Information). 

On top of this novel oscillatory pattern the transconductance of sample A in Fig.~\ref{fig:mapping}b(e) shows faint diagonal lines in the \textit{nn'n}(\textit{pp'p}) regime, which are Shubnikov-de Haas oscillations in the dual-gated region. The occurrence of Shubnikov-de Haas oscillations shows that in this moderate magnetic field regime the Landau levels are broadened by disorder on the scale of their spacing, resulting in a modulation of the density of states.

Using a plate capacitor model described in the supplemental material of Ref.~\citenum{Varlet2014}, the gate voltage axes can be converted into density and filling factor axes, $\nu_X$ with $X = \mathrm{SG},\mathrm{DG}$ for the single- and dual-gated regions, respectively. The result of this transformation is shown in Fig.~\ref{fig:mapping}c. The oscillatory pattern has a slope of one, i.e. it follows lines of constant filling factor difference $\Delta \nu = \nu_\mathrm{SG}-\nu_\mathrm{DG}$. It appears that the oscillations can be described by:
\begin{equation}
G = \langle G \rangle + A ~ \cos(2 \pi \frac{\Delta \nu}{8})
\label{eq:cond}
\end{equation}
where $A$ is the amplitude of the oscillations, which is on the order of 4 \% of the background conductance $\langle G \rangle$ at $T = 1.7$~K. The distance between one conductance maximum and the next can therefore be bridged by either changing the filling factor in one region by 8, or by changing the filling factor in both regions oppositely by 4. It should be noted that Eq.~(\ref{eq:cond}) can be used to describe the oscillations in all six samples, regardless of the number of graphene layers and the sample width (see table \ref{tab:sizes} and the Supporting Information). 

The oscillations persist in magnetic fields up to $B = 1$~T for sample A and the periodicity scales with $\Delta \nu$ for the entire magnetic field range. In higher magnetic fields the conductance is dominated by quantum Hall edge channels and takes on values below $e^2/h$ in the \textit{npn} and \textit{pnp} regimes, in agreement with observations by Amet et al.\cite{Amet2014}. Other works report on the (partial) equilibration of edge channels\cite{Williams2007,Ozyilmaz2007,Ki2010,Amet2014,Morikawa2015,Tovari2016b} and shot noise \cite{Kumada2015,Matsuo2015} in \textit{p}-\textit{n} junctions in the quantum Hall regime.

\begin{figure}
\centering
\includegraphics[width=0.5\textwidth]{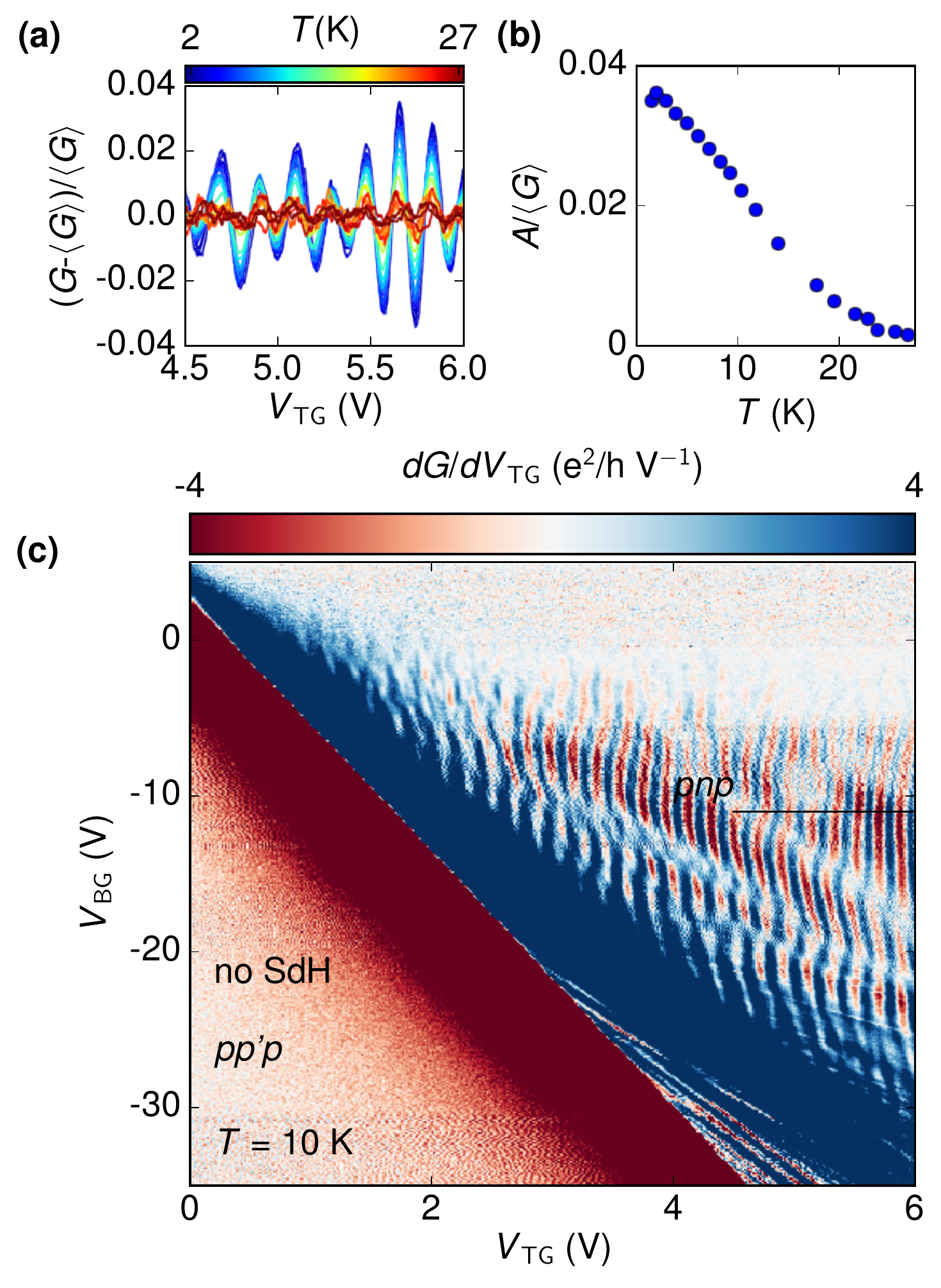}
\caption{Temperature dependence. (a) Oscillatory part of the conductance as a function of top gate voltage and temperature measured along the line cut indicated by the black line in Fig.~\ref{fig:3}c. (b) Amplitude $A$ of the oscillatory conductance as a function of temperature. The oscillations disappear around $T = 30$ K. (c) Transconductance at $T = 10$~K, $B = 0.5$~T in the \textit{pnp} regime. Whereas Shubnikov-de Haas oscillations in the \textit{pp'p} regime have faded out, the oscillatory pattern in the \textit{pnp} regime persists.}
\label{fig:3}
\end{figure}

The oscillatory conductance is quite robust against temperature changes. Figure~\ref{fig:3}a,b show the decay of the amplitude as a function of temperature $T$. The oscillatory conductance in the \textit{pnp} and \textit{npn} regime disappear at a temperature around $T = 30~$K. As can be seen in Fig.~\ref{fig:3}c, at $T = 10$~K the oscillations are still clearly present, while the Shubnikov-de Haas oscillations in the \textit{pp'p} regime have already faded out. The persistence up to $T = 30~$K indicates that the studied phenomenon does not require phase coherence on the scale of the device size. The phase coherence length at $T = 1.7~$K is estimated to be on the order of the device size, but it falls off with $1/T$.\cite{Engels2014}

\begin{figure*}
\centering
\includegraphics[width=1\textwidth]{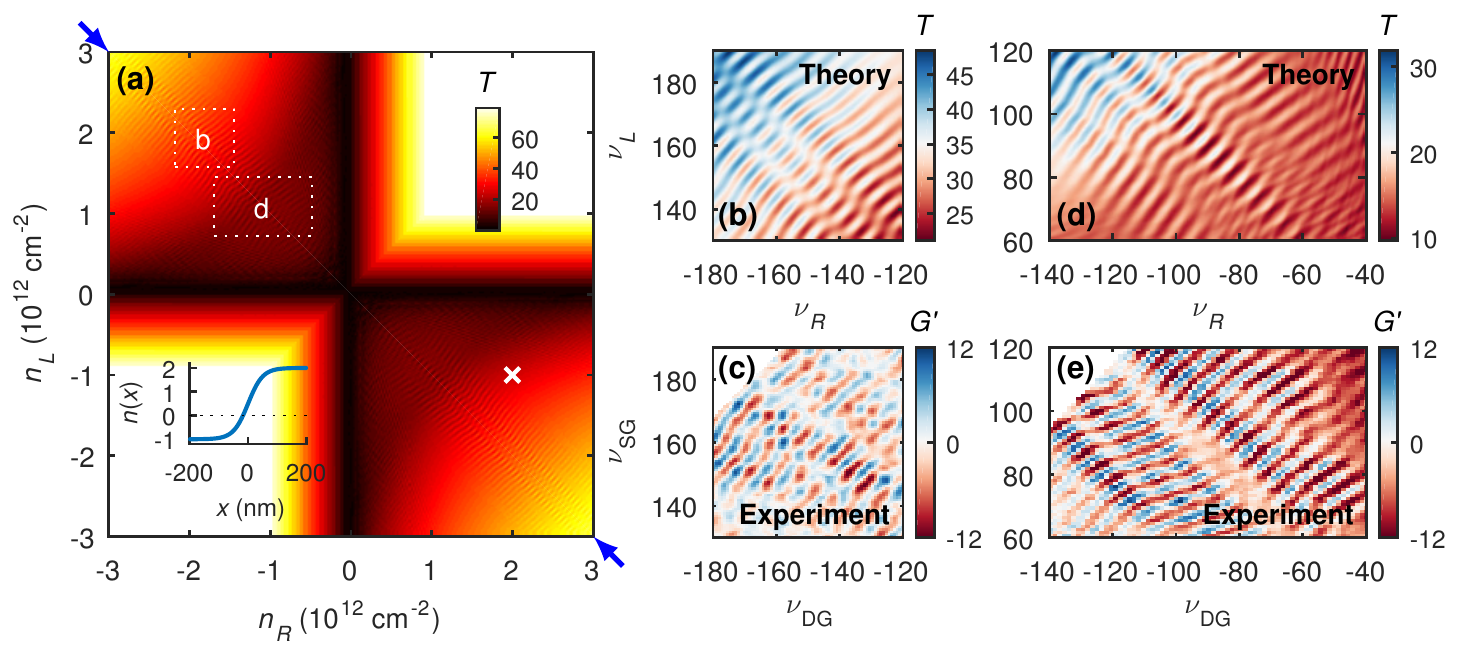}
\caption{(a) Transmission $T$ as a function of the carrier densities on the left, $n_L$, and right, $n_R$, for an ideal SLG \textit{p}-\textit{n} junction at a perpendicular magnetic field $B=0.5$~T based on a tight-binding transport calculation (color range restricted for clarity). Oscillations occur in the vicinity of the symmetric bipolar axis marked by blue arrows. Inset: an example of the considered carrier density profile corresponding to the white cross. White dashed boxes correspond to the density regions shown in panels (b) and (d), where the carrier density values are transformed in filling factors. (c)/(e) Transconductance $G'$ measured for sample B/E shown with the same filling factor range as (b)/(d).}
\label{fig:tb}
\end{figure*}

The above discussed oscillations can be reproduced by transport calculations for an ideal SLG \textit{p}-\textit{n} junction at an intermediate magnetic field $B$, based on the scalable tight-binding model \cite{Liu2015}. The ideal junction is modeled by connecting two semi-infinite graphene ribbons (oriented along armchair) with their carrier densities given by $n_L$ in the far left and $n_R$ in the far right. A simple hyperbolic tangent function with smoothness $50$~nm bridging $n_L$ and $n_R$ is considered; see the inset of Fig.~\ref{fig:tb}a for an example. To cover the density range up to $\pm 3\times 10^{12}$~cm$^{-2}$ corresponding to a maximal Fermi energy of $E_{\max}\approx 0.2$~eV, the scaling factor $s_f=10$ is chosen because it fulfills the scaling criterion \cite{Liu2015} $s_f\ll 3 t_0 \pi/E_{\max}\approx 141$ very well; here $t_0\approx 3$~eV is the hopping energy of the unscaled graphene lattice. Note that the following simulations consider $W=1~\mu$m for the width of the graphene ribbon, but simulations based on a different width show an identical oscillation behavior (see Supporting Information for details), confirming its width-independent nature as already concluded from our measurements.

The transmission function $T(n_R,n_L)$ across the ideal \textit{p}-\textit{n} junction at $B=0.5$~T is shown in Fig.~\ref{fig:tb}a, where fine oscillations along symmetric bipolar axis (marked by the blue arrows) from \textit{np} to \textit{pn} through the global charge neutrality point can be seen. Two regions marked by the white dashed boxes in Fig.~\ref{fig:tb}a are zoomed-in and shown in Figs.~\ref{fig:tb}b and d for a closer look and comparison with the measurements of sample B and E (Figs.~\ref{fig:tb}c and e,  respectively).  Despite certain phase shifts (observed in Figs. 5b, d, and e) that are beyond the scope of the present study, good agreement between our transport simulation and experiment showing the oscillation period well fulfilling Eq.~(\ref{eq:cond}) can be seen.

Other works \cite{Taychatanapat2015,Rickhaus} report on the formation of so called snake states along \textit{p}-\textit{n} interfaces in graphene. Snake states result in a minimum in the conductance whenever the sample width $W$ and the cyclotron radius $R_\mathrm{c}$ satisfy $W/R_\mathrm{c} = 4m-1$ with  $m$ a positive integer.  In the density range of Fig.~\ref{fig:mapping}b,e this would lead to two resonances at most (indicated by green dashed lines in Fig.~\ref{fig:mapping}b,e), which is far less than the observed number of resonances. On top of that, snake states are inconsistent with the observed absence of a dependence on sample width. Furthermore, the tight-binding simulation also confirms that the observed effect is independent of the sample width and cannot be suppressed by introducing strong lattice defects in the vicinity of the \textit{p}-\textit{n} junction (see Supporting Information). We therefore rule out snaking trajectories as a possible cause of the observed oscillations.

Another process which could give rise to oscillations in a graphene \textit{p}-\textit{n} junction in a magnetic field is the interference of charge carriers which are partly reflected and partly transmitted at the interface. When the charge carrier densities are equal on both sides of the interface, electrons and holes will have equal cyclotron radii and therefore the paths of transmitted and reflected charge carriers will form closed loops. For the case of equal density, this model predicts the right periodicity of the oscillations. \cite{Patel2012} Experimentally, however, the measured oscillations are still visible when the densities on both sides of the \textit{p}-\textit{n} interface are quite different: at the point ($V_\mathrm{BG}$,$V_\mathrm{TG}$) = (12,-6)~V  for example (see Fig.~\ref{fig:mapping}b), the cyclotron radii on the \textit{p} and \textit{n} side are respectively 0.36~$\mu$m and 0.16~$\mu$m. The path lengths hence differ by $2 \Delta R_\mathrm{c} = 0.40~\mu$m, which is more than seven times the Fermi wavelength (0.02~$\mu$m and 0.05~$\mu$m). It seems unlikely that interference between charge carriers on skipping orbits can still occur in this density regime. On top of that, the tight-binding simulations show that the oscillatory pattern is still present when introducing large-area lattice defects in the vicinity of the \textit{p}-\textit{n} junction, which destroy the skipping trajectories (see Supporting Information). The observed robustness against temperature changes is in contradiction with this model as well. Thus, the observed oscillations cannot be ascribed to interference of charge carriers on cyclotron orbits at the \textit{p}-\textit{n} interface.

Since the oscillations occur in both single-layer and bilayer graphene, we exclude an explanation that relies on specific details of the dispersion relation. In the magnetic field range where the oscillations are observed, the sample width is comparable to the classical cyclotron diameter. This excludes explanations based on classical electron flow following skipping orbit-like motion along edges.

\begin{figure}
\centering
\includegraphics[width=0.5\textwidth]{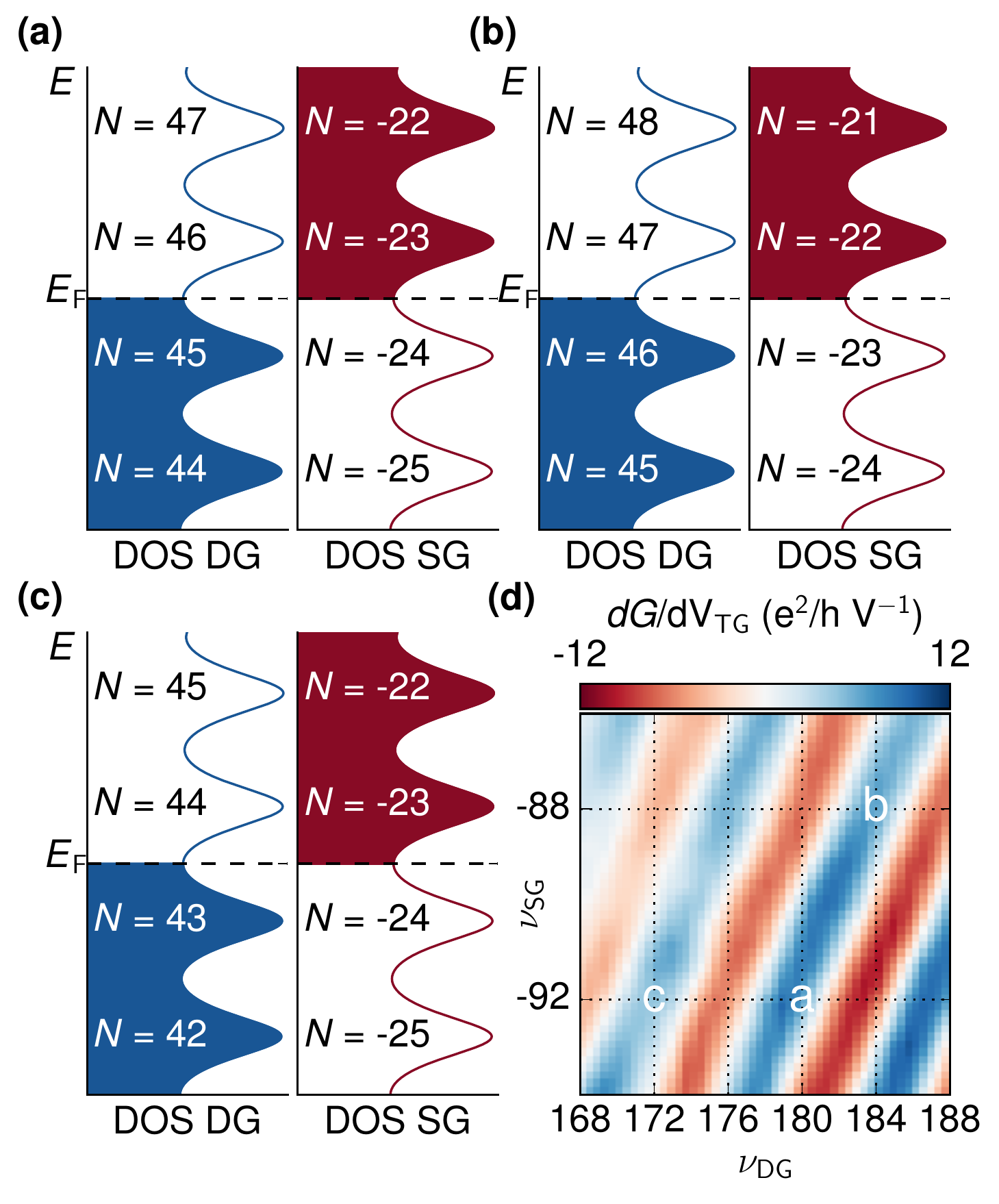}
\caption{(a-c) Schematics of the densities of states as a function of energy for the single- and dual-gated regions of the sample at positions a-c of the measurement (d) Zoom in Fig.~\ref{fig:mapping}c. The alignment of the densities of states can lead to an oscillatory pattern with the right slope: along the line from point a to point b the two combs of densities of states stay aligned, whereas the combs shift with respect to one another when moving from point a to point c.}
\label{fig:4}
\end{figure}

A mechanism which may cause the oscillations involves the alignment of the density of states (DOS) around the Fermi energy. Diagrams of the DOS in the single- and dual-gated regions are sketched in Figs.~\ref{fig:4}a-c. Figure~\ref{fig:4}d shows a zoom in the map of the oscillatory  transconductance of Fig.~\ref{fig:mapping}c. At point \textsf{a} in this zoom the filling factor in the dual-gated regime is $\nu_\mathrm{DG} =$~180 and $\nu_\mathrm{SG} =$~-92 in the single-gated region. Because of the fourfold degeneracy of the Landau levels, Landau level numbers are $N =$~45 and $N =$~-23 respectively, as shown in the DOS diagram of Fig.~\ref{fig:4}a. When following the oscillatory pattern from point \textsf{a} to point \textsf{b}, the two combs of DOS remain aligned with one another and only the Fermi level changes. This in contrast to what happens when moving from point \textsf{a} to point \textsf{c}: the DOSs shift with respect to one another and the transconductance oscillates. It could therefore be the case that the alignment of the DOS affects the conductance of the \textit{p}-\textit{n} interface in a way similar to the magneto-intersubband oscillations (MISO) of a two-dimensional electron gas (2DEG):\cite{Raikh1994,Dmitriev2012} the occupation of two energy subbands of a 2DEG can lead to enhanced scattering between the subbands when the DOSs of the subbands are aligned. Although the \textit{p}- and \textit{n}- regions are spatially separated in the case of graphene \textit{p}-\textit{n} junctions, a similar enhancement of the coupling at the interface may be observed.  In the \textit{pp'p} and \textit{nn'n} regime the interfaces are much more transparent (see conductance in Fig.~\ref{fig:3}a,d and Ref.\citenum{Rickhaus2015}), therefore the interface plays a negligible role in the total conductance. This explains why the oscillations are only visible in presence of a \textit{p}-\textit{n} interface. As the two outer regions of the sample have the same density up to an insignificant difference in residual doping, the two interfaces contribute in a similar way. The number of interfaces can at most influence the visibility of the oscillations. In practice we find that the visibility is however mostly influenced by sample quality. Just as for the oscillations we report on, MISO persist up to relatively high temperatures. The spacing predicted by this model lacks a factor of two compared to the experiment, however: it would predict the argument of the cosine of Eq.~(\ref{eq:cond}) to be $2 \pi \Delta \nu / 4$. Further investigation is needed to explain this discrepancy between the MISO model on the one hand and the experimental data and the tight-binding simulations on the other hand.

In conclusion, we have observed oscillations in the conductance of six graphene \textit{p}-\textit{n} junctions in the magnetic field range of $B = 0.4  - 1.5$~T. The oscillations are independent of sample width and can be described by the filling factor difference between the single- and dual-gated regions. The oscillations are quite robust against temperature changes: they fade out only in the range of $T = 20 - 40$~K, whereas Shubnikov-de Haas oscillations decay below $T = 10$~K. The oscillations can be well reproduced by tight-binding transport calculations considering an ideal p-n junction at a constant magnetic field. Up to a factor of two, the oscillatory pattern can be explained by considering the density of states alignment of the single- and dual-gated regions.

\section*{acknowledgement}
We thank Péter Makk and François Peeters for fruitful discussions. We acknowledge financial support from the European Graphene Flagship and the Swiss National Science Foundation via NCCR Quantum Science and Technology. M.-H.L. and K.R. acknowledge financial support by the Deutsche Forschungsgemeinschaft within SFB 689.

\newpage
\bibliographystyle{apsrev4-1}
\begingroup
\inputencoding{latin1}
\bibliography{oscillations}
\endgroup
\clearpage


\section*{Supplementary material}

# Supporting Information for
# Oscillating Magnetoresistance in Graphene p-n Junctions at Intermediate Magnetic Felds

Hiske Overweg,[1] Hannah Eggimann,[1] Ming-Hao Liu,[2,3] Anastasia Varlet,[1] Marius Eich,[1] Pauline Simonet,[1] Yongjin Lee,[1] Kenji Watanabe,[4] Takashi Taniguchi,[4] Klaus Richter,[2] Vladimir I. Fal'ko,[5] Klaus Ensslin,[1] and Thomas Ihn[1]

[1]*Solid State Physics Laboratory, ETH Zürich, CH-8093 Zürich, Switzerland*[*]
[2]*Institut für Theoretische Physik, Universität Regensburg, D-93040 Regensburg, Germany*
[3]*Department of Physics, National Cheng Kung University, Tainan 70101, Taiwan*
[4]*National Institute for Material Science, 1-1 Namiki, Tsukuba 305-0044, Japan*
[5]*National Graphene Institute, University of Manchester, Manchester M13 9PL, UK*

## TIGHT BINDING SIMULATIONS

### Overview

In the main text, we have shown a transmission map as a function of left and right carrier density, calculated using the real-space Green's function method based on the scalable tight-binding model [1]. The considered graphene ribbon of width $W = 1\,\mu m$ is subject to a model density function describing an ideal *pn* junction with smoothness 50 nm. The full map is repeated here in Fig. S1(a), with a white box marking the region plotted in Fig. S1(b).

In this Supporting Information, we show more numerical results in order to demonstrate that the observed oscillation is independent of the smoothness of the *pn* junction and the width of the graphene ribbon, and is not related the current along the *pn* junction. Instead, the oscillation is shown by the last numerical test to be closely related to the Landau levels away from the *pn* junction.

For quantitative and systematic comparisons, we will focus on the density range shown in Fig. S1(b) and the line cut on it along the dashed line shown in Fig. S1(c). All calculations shown in the following consider the same density range and resolution as Figs. S1(b) and (c), which can be regarded as the reference panels of this Supporting Information. In particular, the line cut of Fig. S1(c) will be repeatedly shown in the following results.

### Smoothness dependence

Figure S2(a) presents the transmission map with smoothness of 25 nm, showing a similar pattern seen in Fig. S1(b) where the junction smoothness is 50 nm. A more quantitative comparison is shown in Fig. S2(b) for the line cuts of the two cases. Despite a slightly higher $T$ obtained for the sharper junction due to the Klein collimation [2], i.e., the sharper the *pn* junction, the wider the finite transmission probability of the angle distribution and hence the conductance, the general trend of the oscillation is shown to be independent of the smoothness.

In the rest of the numerical results, the smoothness will be fixed to 50 nm.

### Width dependence

Figure S3 presents the transmission map based on a graphene ribbon with $W = 0.9\,\mu m$ shown in its panel (a) and $W = 0.8\,\mu m$ shown in its panel (b). Comparing the line cuts in Fig. S3(c), along with the reference line of Fig. S1(c) for the case of $W = 1\,\mu m$, the feature of the oscillation is clearly shown to be width-independent. On the other hand, the oscillation amplitude decreasing with the reduced graphene width implies that the oscillation may be closely related to the Landau levels in the bulk away from the *pn* junction, since the wider the graphene ribbon the better the Landau levels can develop.

In the rest of the numerical results, the graphene width will be fixed as $W = 1\,\mu m$.

### Strong lattice defects

Next we show that the oscillation is not related to the current along the *pn* junction. To this end, we consider large-area lattice defects located in the vicinity of the *pn* junction. The basic idea is that if the oscillation came from any interference due to the current along the *pn* junction, such as the snake state [3], a large-area lattice defect at the *pn* interface or in the vicinity of it would act as a strong scatterer, destroying the

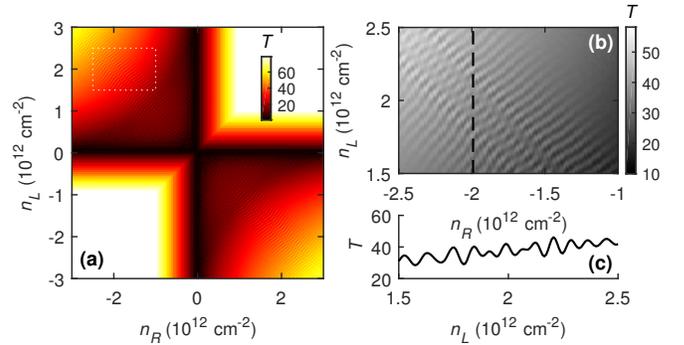

FIG. S1. (a) Transmission map $T(n_R, n_L)$ same as Fig. 5(a) in the main text (smoothness 50 nm and graphene width $W = 1\,\mu m$); white dashed box marks the region shown in (b), where a black dashed line indicates the line cut of $T(n_L)$ at fixed $n_R \approx -2 \times 10^{12}\,cm^{-2}$ shown in (c).

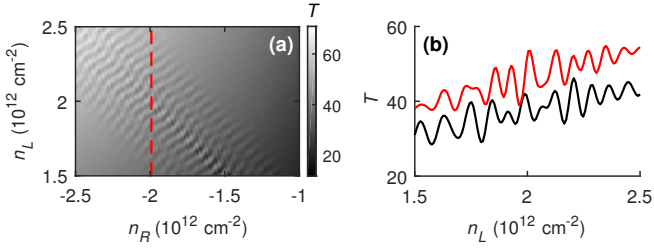

FIG. S2. (a) Transmission map $T(n_R, n_L)$ similar to Fig. S1(b) but with smoothness 25 nm of the $pn$ junction. The red dashed line indicates the line cut shown as a red line in (b), where the black line is the reference line identical to Fig. S1(c) for the case with smoothness 50 nm.

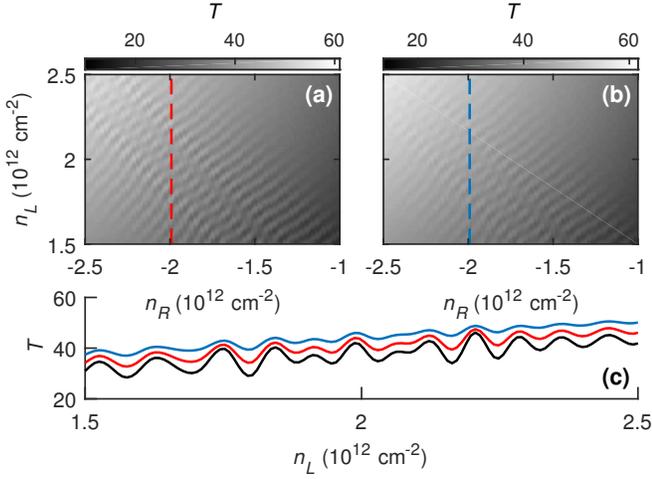

FIG. S3. Transmission maps $T(n_R, n_L)$ similar to Fig. S1(b) with the same smoothness of 50 nm but with (a) $W = 0.9\,\mu$m and (b) $W = 0.8\,\mu$m. Line cuts along the red/blue dashed line marked in (a)/(b) are compared in (c) together with the reference line (black) of Fig. S1(c) for the case of $W = 1\,\mu$m.

interference and hence suppressing the oscillation. Contrarily, if the oscillation survives the introduced large defects, the current along the $pn$ junction will then be ruled out from possible origins of the oscillation.

We first consider a $50 \times 400\,$nm$^2$ defect in Figs. S4(a) and (b); the defect is placed in front of the $pn$ junction (at a distance 150 nm) in the former, and exactly on the $pn$ junction in the latter. Despite an additional modulating pattern observed in Fig. S4(a), the fine oscillation patterns remain visible in both cases. By increasing the defect area to $300 \times 300\,$nm$^2$, the transmission map shown in Fig. S4(c) still exhibits the same oscillation pattern. A quantitative comparison of the line cuts summarized in Fig. S4(d) together with the reference line from Fig. S1(c) clearly shows that the oscillations observed in Figs. S4(a)–(c) belong to the same type as all those shown previously.

The fact that the strong defect introduced in the vicinity of the $pn$ junction cannot suppress the oscillation clearly indicates that any possible interference effect due to the current

along the $pn$ junction cannot be the origin causing the oscillation. Instead, the oscillation seems to depend only on the Landau levels that are well developed in the semi-infinite leads.

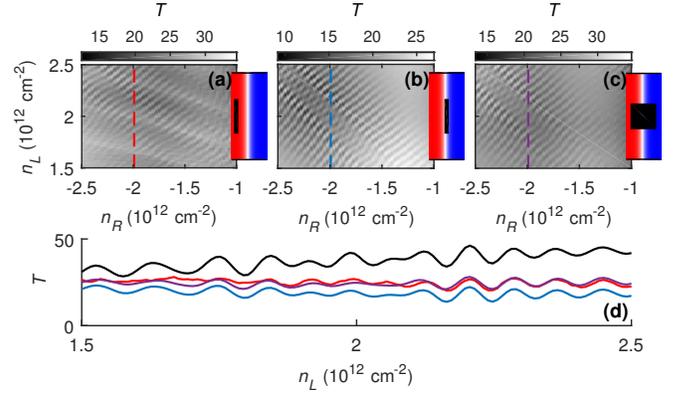

FIG. S4. (a)–(c) Transmission maps $T(n_R, n_L)$ similar to Fig. S1(b) with the same smoothness of 50 nm and width $W = 1\,\mu$m, but with a large-area defect represented by the black rectangle shown in the individual inset to the right of each panel, where the color background depicts the $y$-independent model function $n(x, y)$ describing the density variation of the $pn$ junction. The size of the defect is $50 \times 400\,$nm$^2$ in (a,b) and $300 \times 300\,$nm$^2$ in (c). Line cuts along the red/blue/purple dashed line marked in (a)/(b)/(c) are compared in (d) together with the reference line (black) of Fig. S1(c) for the case without the defect.

### Fixed leads

So far, all the presented calculations are based on an infinite graphene ribbon with a $pn$ junction in the middle, as described in the main text. Technically, this is achieved in numerics by considering a scattering region of size $L \times W$ attached to two leads from the left and right, both floating with the density profiles at the attaching edge of the scattering region. As long as $L$ is much longer than the smoothness of $pn$ junction ($L = 400$ nm has been adopted in all the presented calculations), the density values at the left and right edges of the scattering region will saturate to a constant, and the entire open quantum system of the finite-size scattering region attached to the two floating semi-infinite leads will resemble an ideal $pn$ junction in the middle of an infinitely long graphene ribbon, exhibiting an $L$-independent transmission behavior.

As a final and conclusive numerical test, we now fix the Fermi energies in the two semi-infinite leads at 0.1 eV, and consider the same range and parameters as the reference panel of Fig. S1(b). The calculated transmission map is shown in Fig. S5(a), which no longer exhibits the fine oscillation. The line cuts of fixed leads vs. floating leads compared in Fig. S5(b) clearly show that the oscillation completely vanishes in the present case of fixed leads.

The vanishing oscillation is consistent with what we have speculated from the previously shown tests that the oscillation originates from the resonance between Landau levels well de-

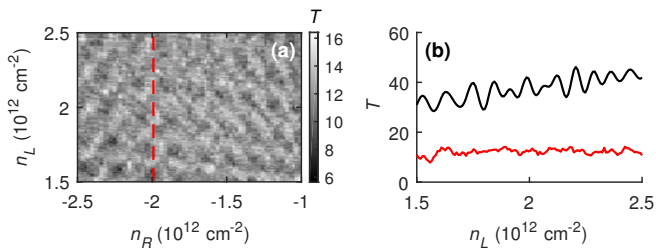

FIG. S5. (a) Transmission map $T(n_R, n_L)$ with the same range and parameters considered in Fig. S1(b) but with the two leads fixed at energy $E = 0.1\,\mathrm{eV}$. The red dashed line indicates the line cut shown as a red line in (b), where the black line is the reference line identical to Fig. S1(c) for the case with floating leads.

veloped in the far left and far right in the semi-infinite leads. The present case shown in Fig. S5 considers fixed Fermi energies in the leads that no longer float with the densities $n_R$ and $n_L$. Together with the fact that the length $L = 400\,\mathrm{nm} \ll W$ of the scattering region is too short for the Landau levels to form, the vanishing oscillation is therefore reasonably expected. By increasing the length of the scattering region to at least $L \approx W$, revival of the oscillation is expected for the case of fixed leads.

Note that the situation of fixed leads is actually closer to the experiment, because the densities in graphene regions close to the contacts are rather pinned by the contact doping. However, the samples in our experiments (summarized in Table I in the main text) are long enough (several microns in all samples) for the Landau levels to develop well (with level spacing not far enough compared to disorder broadening in the magnetic field range we focus on) due to their cleanness and therefore exhibit the oscillation. Our numerical results based on floating leads correspond to the ideal case of infinitely long samples and therefore exhibit optimized oscillation.

## MAGNETORESISTANCE OSCILLATIONS IN SAMPLES B-F

Figures S6-S10 show magnetoresistance oscillations of samples B-F, which look similar to the ones observed in sample A (see Fig. 3 of main text). The periodicity of the oscillations is the same for all samples.

### 4-terminal measurements in sample D

The device layout of sample D is schematically shown in Fig. S11. A DC voltage of $100\,mV$ was applied to the sample with a $R = 10\ \mathrm{M\Omega}$ resistor in series. This led to a constant current of $I = 10\ \mathrm{nA}$ flowing from contact 1 to contact 4. The

voltage drop between contact pairs (1,2), (2,3) and (3,4) where measured. To calculate the conductance a contact resistance was subtracted where appropriate. As is shown in Fig. S11b-d, the oscillatory magnetoresistance is only observed when a $p$-$n$ interface is present (i.e. in between contacts (2,3)).

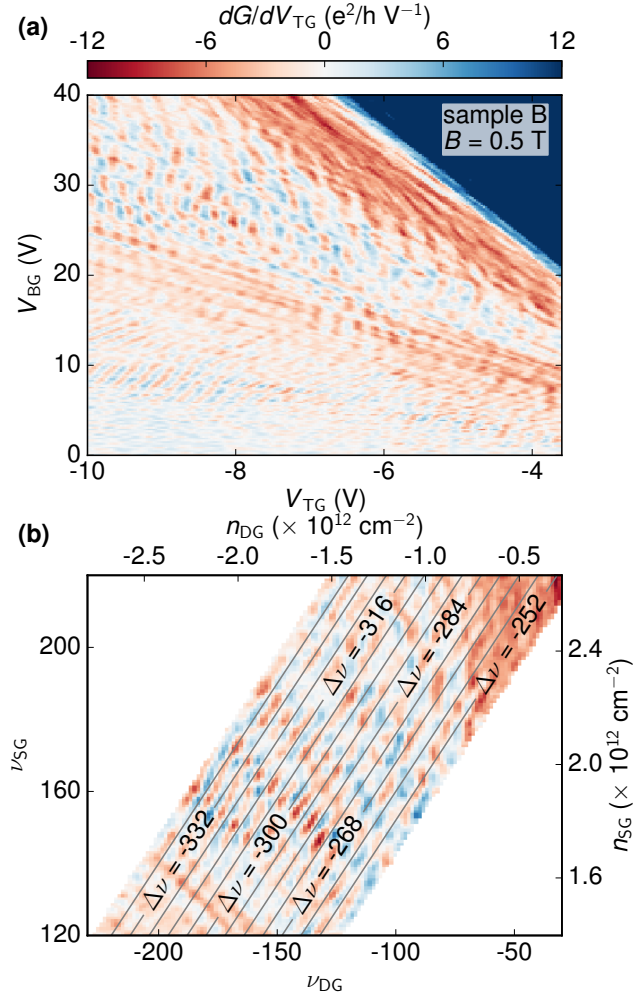

FIG. S6. (a) Transconductance $dG/dV_{\mathrm{TG}}$ of sample B at $B = 0.5$ T. (b) Transconductance as a function of filling factor in the single and double gated region.

* overwegh@phys.ethz.ch

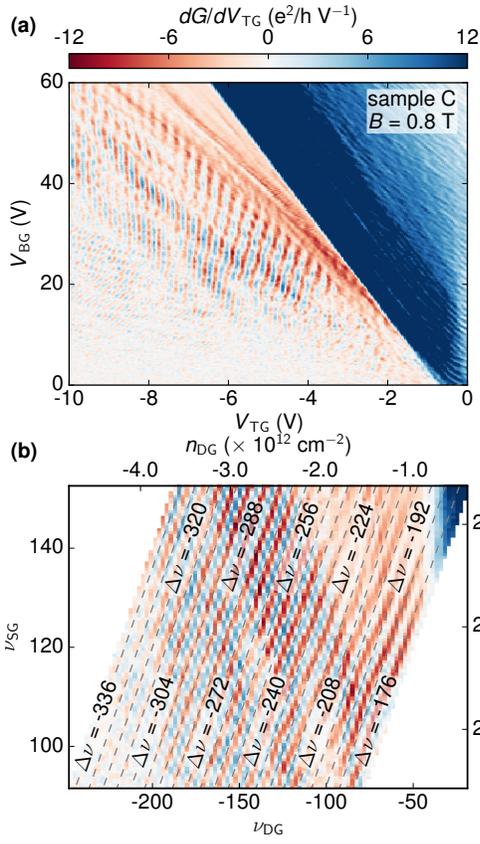

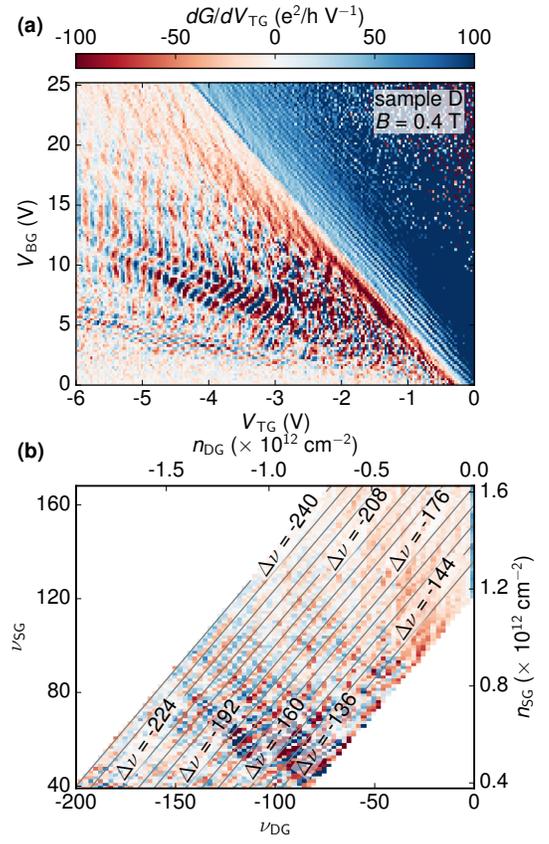

FIG. S7. a) Transconductance $dG/dV_{TG}$ of sample C at $B = 0.8$ T. (b) Transconductance as a function of filling factor in the single and double gated region.

FIG. S8. (a) Transconductance $dG/dV_{TG}$ of sample D at $B = 0.4$ T. (b) Transconductance as a function of filling factor in the single and double gated region.

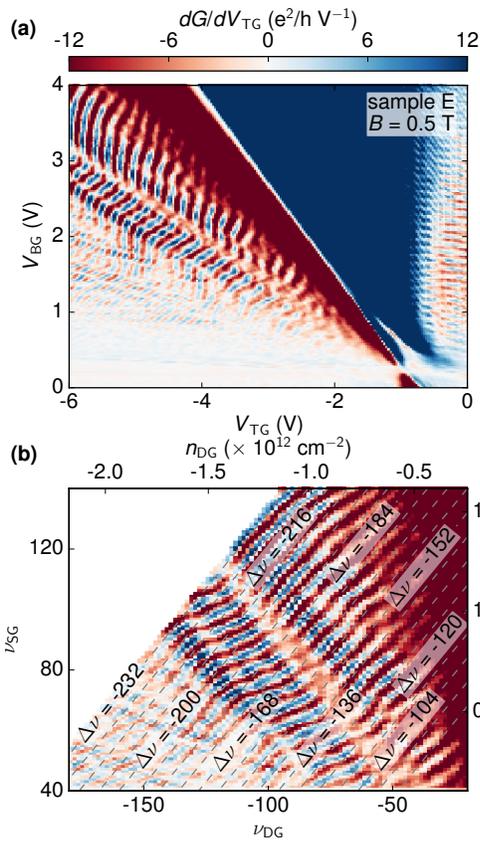

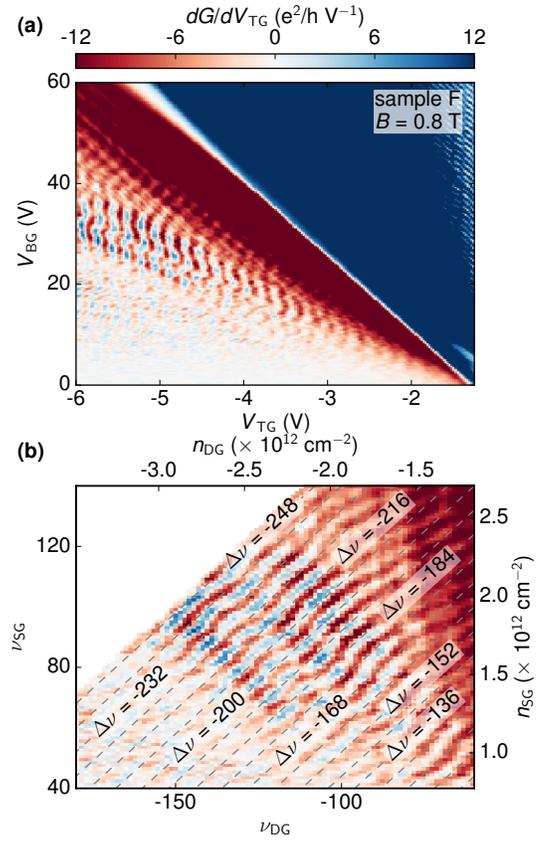

FIG. S9. (a) Transconductance $dG/dV_{TG}$ of sample E at $B = 0.5$ T. (b) Transconductance as a function of filling factor in the single and double gated region.

FIG. S10. (a) Transconductance $dG/dV_{TG}$ of sample F at $B = 0.8$ T. (b) Transconductance as a function of filling factor in the single and double gated region.

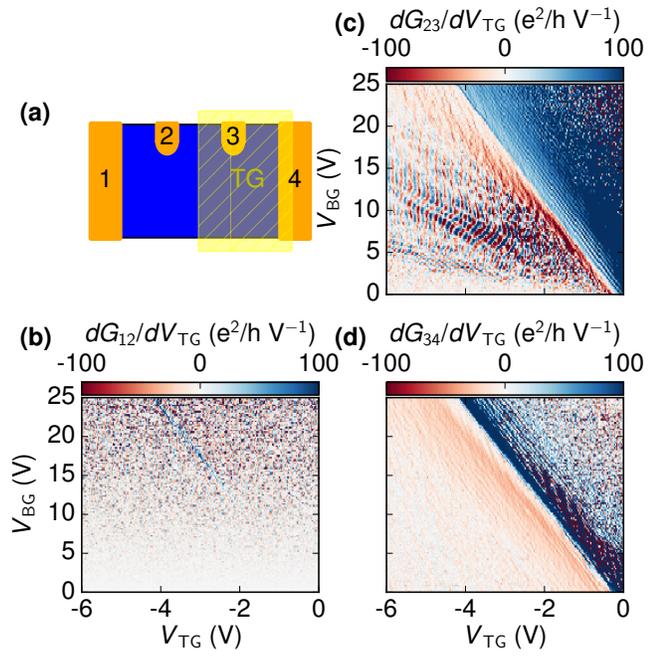

FIG. S11. (a) Schematic drawing of sample D. Contacts are labelled 1-4. (b) Transconductance between contacts (1,2) at $B = 0.4$ T. (c) Transconductance between contacts (2,3) showing an oscillatory pattern in the *p-n* regime. (d) Transconductance between contacts (3,4).


\end{document}